\documentclass[onecollarge,natbib]{svjour2}
\bibpunct{[}{]}{;}{n}{}{,} 
\smartqed  
\usepackage{graphicx}
\usepackage{amsmath}
\usepackage{amssymb}
\usepackage{bm,curves}
\usepackage{xcolor}

\def\IE{{\it i.e.,}}
\def\EG{{\it e.g.,}}

\def\lan{\langle }
\def\ran{\rangle }

\journalname{Few-Body Systems}
\begin{document}

\title{Phenomenological implications of the nucleon's meson cloud
\thanks{This material is based upon work supported by the U.S.~Department of Energy Office of Science,
Office of Basic Energy Sciences program under Award Number DE-FG02-97ER-41014. NT@UW-14-24.}
}

\author{T.~J.~Hobbs
}

\institute{T.~J.~Hobbs \at
              Department of Physics, University of Washington, Seattle, WA 98195-1560 \\
              Tel.: +1-206-543-9754\\
              \email{tjhobbs@uw.edu}           
}

\date{Received: date / Accepted: date}

\maketitle

\begin{abstract}
The long-distance structure of the interacting nucleon receives important contributions from its
couplings to light hadronic degrees of freedom --- a light meson cloud --- while an analogous
nonperturbative mechanism is expected to generate an intrinsic charm (IC) component to the proton wavefunction.
We investigate both possibilities, keeping for the former a special eye to improving the theoretical understanding of the
pion-nucleon vertex in light of proposed measurements. Regarding the latter possibility of IC, we highlight recent results
obtained by a global QCD analysis of the light-front model proposed in Ref.~\cite{Hobbs:2013bia}.
\end{abstract}

\section{Introduction}
\label{sec:intro}

As much as a desire to understand bound state properties of hadrons motivated the development of QCD, efforts based solely
upon perturbative QCD remain stubbornly unyielding toward a thorough grasp of long-range hadronic structure. The defining
reason has much to do with the fact that nonperturbative mechanisms are decisive in shaping various aspects of 
the makeup and internal dynamics of hadrons. That this must be the case is evident, for instance, in the role played by
light pionic modes in qualitatively shaping the peripheral charge distribution of the nucleon due to pion-nucleon loop
corrections at its electromagnetic vertex. Such corrections are a direct consequence of the nucleon's {\it pion cloud}
--- essentially, a dressing of the nucleon wavefunction by short-lived configurations of virtual $SU(2)$ mesons, of which the
pion as the lightest mode is expected to dominate at small momenta ($\lesssim 500$ MeV). In fact, at small values of the
$t$-channel exchange mass ($t \sim m^2_\pi$) the presence of the pion cloud provides a natural description of the electron-nucleon
DIS interaction in terms of spontaneous dissociations of the form $N \rightarrow \pi N$ under the constrains of, {\it e.g.}, charge
conservation and gauge invariance.

The existence of the pion cloud is in principle well-established
by a substantial body of experimental measurements and theoretical analyses ({\it e.g.}, of light quark sea asymmetries
$[\bar{u} - \bar{d}](x) \neq 0$ relevant to Gottfried Sum Rule violation \cite{Tony}). At the same time, the detailed
momentum dependence of the pion-nucleon vertex remains a significant source of model dependence in any effort at
first-principles calculation; proposed DIS extractions of the pion structure function (SF) $F^\pi_2(x,Q^2)$ at the crucial
pion mass pole $t = m^2_\pi$ depend upon sensitivity to mechanisms including those depicted in Fig.~\ref{fig:Feyn} for
which systematic estimations at the relevant kinematics remain lacking --- an issue we explore in Sec.~\ref{sec:TDIS}
in light of the experimental possibility of forward spectator tagging \cite{TDIS}. On the other hand, though the validity of
the ``cloud'' picture might break down with increasing mass, the basic meson-baryon Fock state ansatz also remains a sensible means of
determining nonperturbative contributions from heavy quarks \cite{Hobbs:2013bia}, as we highlight with a
summary of a related QCD global fit in Sec.~\ref{sec:GA}.

\section{Pion-nucleon interactions and spectator tagging}
\label{sec:TDIS}

\begin{figure*}
\centering
\includegraphics[width=0.32\textwidth]{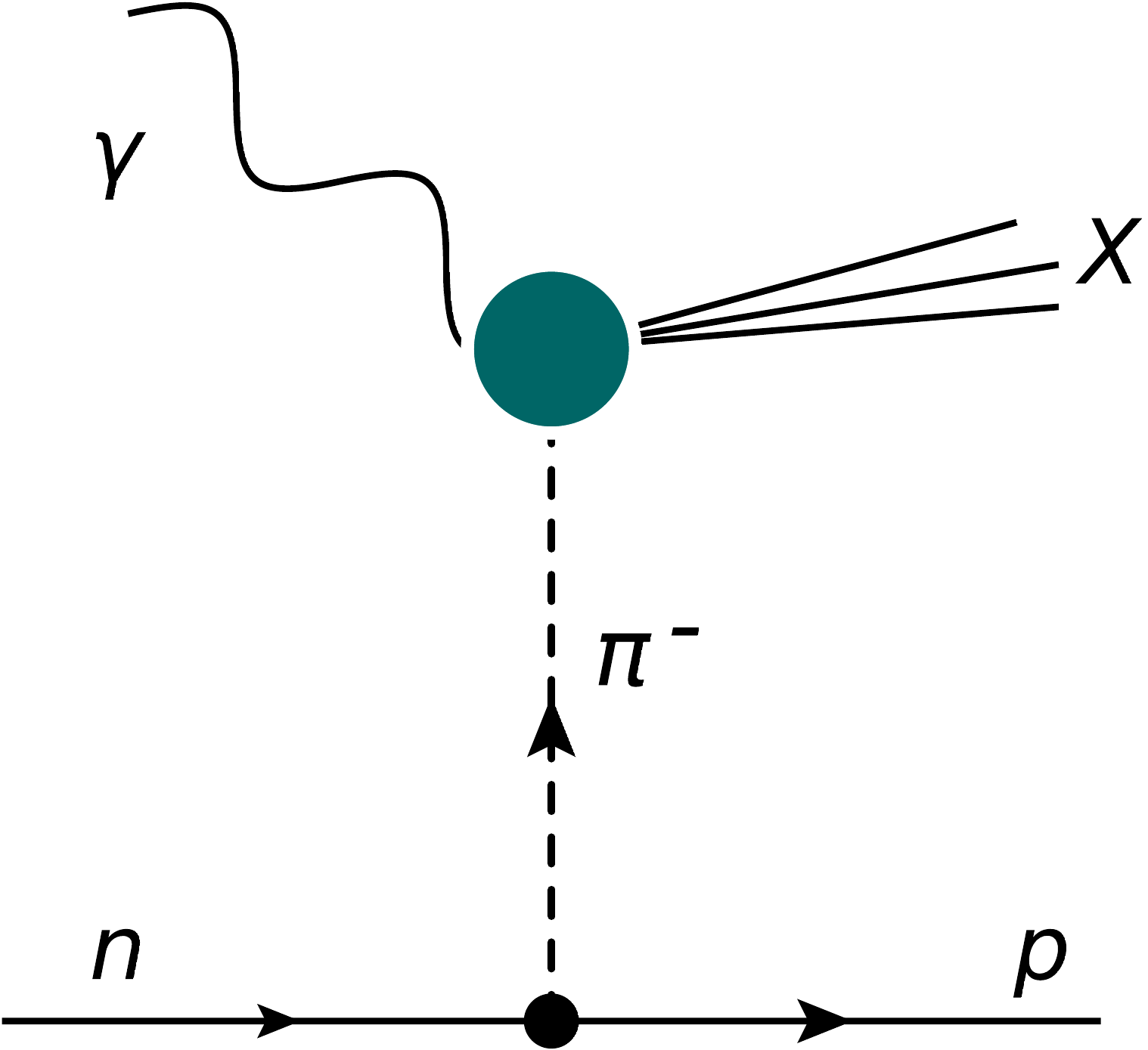} \ \ \ \ \ \ \ \
\includegraphics[width=0.37\textwidth]{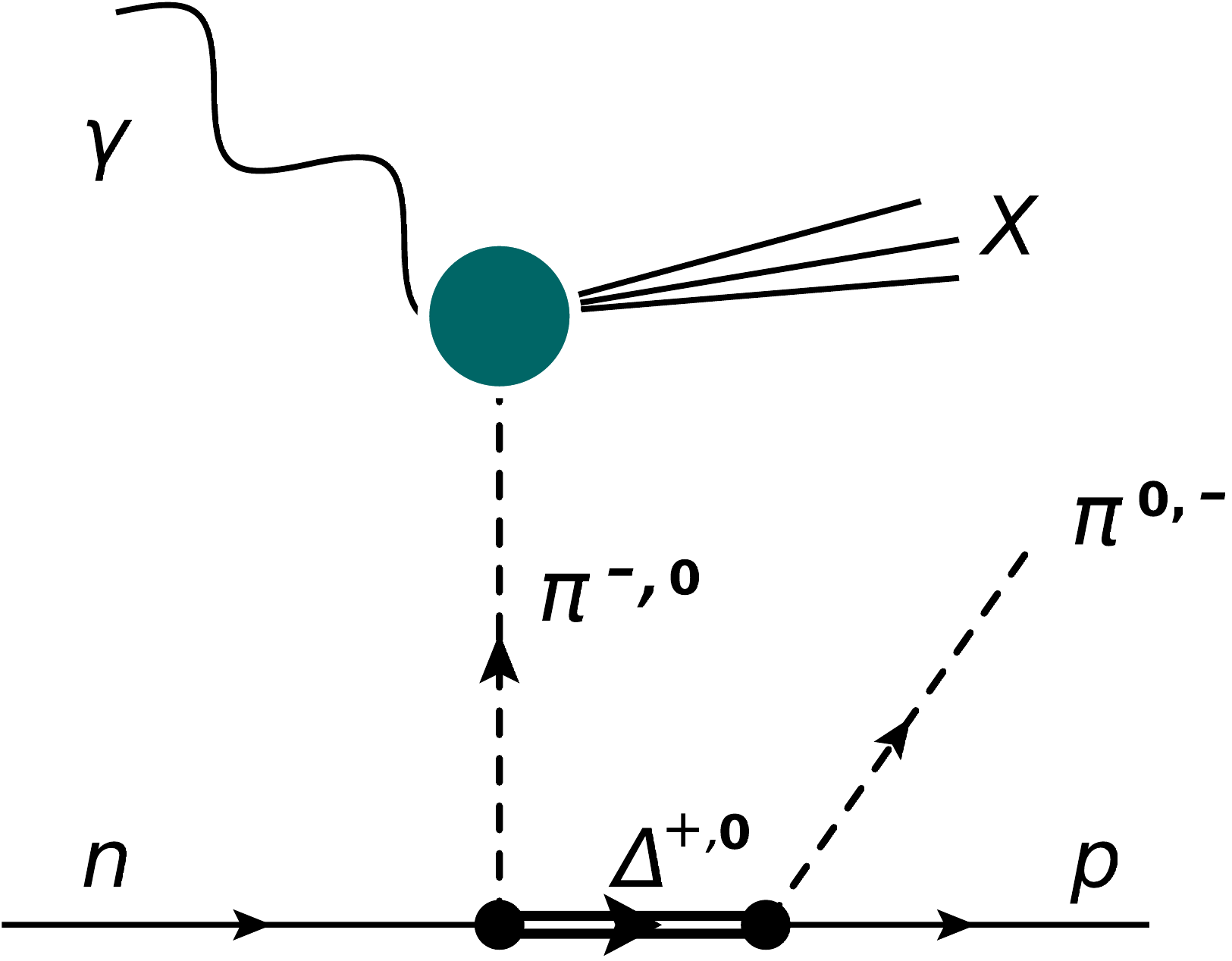}
\caption{We take the dominant process to be light meson exchange
as typified by these one-pion graphs; the effects of intermediate $\Delta$
states are also considered numerically.}
\label{fig:Feyn}
\end{figure*}

To lowest order, the process of Fig.~\ref{fig:Feyn} implies a two-component Fock state expansion of the type
\begin{equation}
|N\rangle\ =\ \sqrt{Z_2}\, \left| N \right.\rangle_0\
+\ \sum_{M,B} \int\! dz\, f_{MB}(z)\, | M(z); B(1-z) \rangle\ ,
\label{eq:Fock}
\end{equation}
where for full generality, the wavefunction of the physical nucleon is taken to be a direct
sum of a renormalized bare term with a contribution from meson-baryon virtual states. The
meson-baryon splitting function $f_{MB}(z) = \int dk^2_\perp\ \phi_{MB}(z, k^2_\perp)$ requires
the computation of the amplitudes $\phi_{MB}(z, k^2_\perp)$, which we undertake on the light-front
in terms of the momentum fraction $z = k^+ / P^+$ of the intermediate meson with respect to
the initial state nucleon. Fortunately, $\chi$PT provides the necessary lagrangians to satisfy
gauge invariance, and for the pion sector we take the relevant interaction to be
\begin{eqnarray}
{\cal L}_{\pi N}
&=& {g_A \over 2 f_\pi}\,
    \bar\psi_N \gamma^\mu \gamma_5\,
        \bm{\tau} \cdot \partial_\mu \bm{\pi}\, \psi_N\
 -\ {1 \over (2 f_\pi)^2}\,
    \bar\psi_N \gamma^\mu\, \bm{\tau} \cdot
        (\bm{\pi} \times \partial_\mu \bm{\pi})\, \psi_N\ ,
\label{eq:piN-PV}
\end{eqnarray}
in which the leading term in Eq.~(\ref{eq:piN-PV}) induces characteristic ``rainbow'' graphs whence the
pionic $\phi_{MB}(z, k^2_\perp)$ may be derived, and the second, Weinberg-Tomozawa term does not contribute to the
present analysis. Following a standard process for the pseudoscalar pion, one may obtain
\begin{equation}
f_{\pi N}(z)
= c_I \frac{g_{\pi NN}^2}{16 \pi^2}
  \int_0^{\infty} \frac{dk^2_\perp}{(1-z)}
  \frac{G_{\pi N}^2}{z \ (M^2 - s_{\pi N})^2}
  \left( \frac{k^2_\perp + z^2 M^2}{1-z} \right)\ ,
\label{eq:piN-SF}
\end{equation}
where $k_\perp$ is the transverse momentum of the exchanged pion in the Sullivan process of
Fig.~\ref{fig:Feyn}, $g_{\pi NN}$ is the well-known $\pi NN$ coupling constant, $M$ the nucleon mass, and the
isospin factors for the individual charge states are $c_I=1$ for $\pi^0$ ($p \to p \pi^0$ or $n \to n \pi^0$)
and $c_I=2$ for $\pi^\pm$ ($p \to n \pi^+$ or $n \to p \pi^-$). While the spin-structure
and general form of Eq.~(\ref{eq:piN-SF}) are fully determined diagrammatically, the UV-divergent
behavior that originates with the implicit $k^2_\perp$ integration of Eq.~(\ref{eq:Fock}) is a primary
source of model dependence, with a variety of phenomenological form factors traditionally used
to control the irregular behavior at large momenta; here we make use of an $s$-dependent scheme:
$G_{\pi N} = \exp\left[(M^2 - s_{\pi N})/\Lambda^2\right]$.

With these, as well as corresponding expressions for other $SU(2)$ processes involving scattering
from the $\rho$ or intermediate $\Delta$ exchanges, etc., we are in a position to make definite
predictions for experimental observables, including meson cloud components of DIS structure functions.
In keeping with the current model uncertainties regarding the pion-nucleon vertex described at the outset
of the present note, it would be particularly helpful for experiments to measure the momentum dependence of contributions
from the processes depicted in Fig.~\ref{fig:Feyn}. In practice, this might best be achieved via an experimental
program to systematically ``tag'' produced baryons as well as carefully measure their final-state momentum with
sufficiently sensitive forward calorimetry. With the resulting information, one could then deduce on the basis of momentum
conservation the kinematics of the one-boson exchanges shown in Fig.~\ref{fig:Feyn}, and thereby constrain models
of the type specified by Eq.~(\ref{eq:piN-SF}).

Thus, with explicit calculations for the pion-nucleon distribution amplitudes, (\EG~Eq.~[\ref{eq:piN-SF}]), we
can write the fully inclusive contribution of the pion sector to the DIS structure function of the nucleon: 
\begin{equation}
F_2^{(\pi N)}(x)
= \int_x^1 dz\, f_{\pi N}(z)\ F_{2\pi}\Big(\frac{x}{z}\Big)\ ;
\label{eq:conv}
\end{equation}
we will find it useful to recast this in terms of lab-frame variables,
\begin{align}
\tilde{z} = \Big( &k_0 + |{\bm k}| \cos\theta_{p'} \Big) \Big/ M\ , \hspace*{2.5cm} \tilde{k}_\perp = |{\bm k}| \sin\theta_{p'}  \nonumber\\ 
&k_0 = M - \sqrt{M^2+{\bm k}^2}\ ,
\label{eq:kin+}
\end{align}
where $|{\bm k}|$ is the deposited three-momentum of the final-state baryon, unobserved in the inclusive process
implied by Eq.~(\ref{eq:conv}), and $\theta_{p'}$ is its laboratory angle relative to the longitudinal direction defined
by an incident photon. In contrast to these inclusive considerations, a detailed mapping of the momentum dependence of one-pion exchange in
DIS requires a careful binning of the contributions embodied by Eq.~(\ref{eq:conv}) in finite ranges of the transverse
momentum $k^2_\perp \in \Delta k^2_\perp $ and fraction $z \in \Delta z$ (using the ``natural'' parametrization of the pion
cloud). The resulting ``tagged structure functions'' are then specified as
\begin{equation}
F_2^{(\pi N)}(x,\Delta z,\Delta k_\perp^2)
= {1\over M^2} \int_{\Delta z} dz \int_{\Delta k_\perp^2} dk^2_\perp\
  f_{\pi N}(z,k_\perp)\ F_{2\pi}\Big(\frac{x}{z}\Big)\ ,
\label{eq:sidis_int}
\end{equation}
in which the factor of $M^{-2}$ is present to preserve the dimensionlessness of the integral.

\begin{figure*}
\centering
\includegraphics[width=0.48\textwidth]{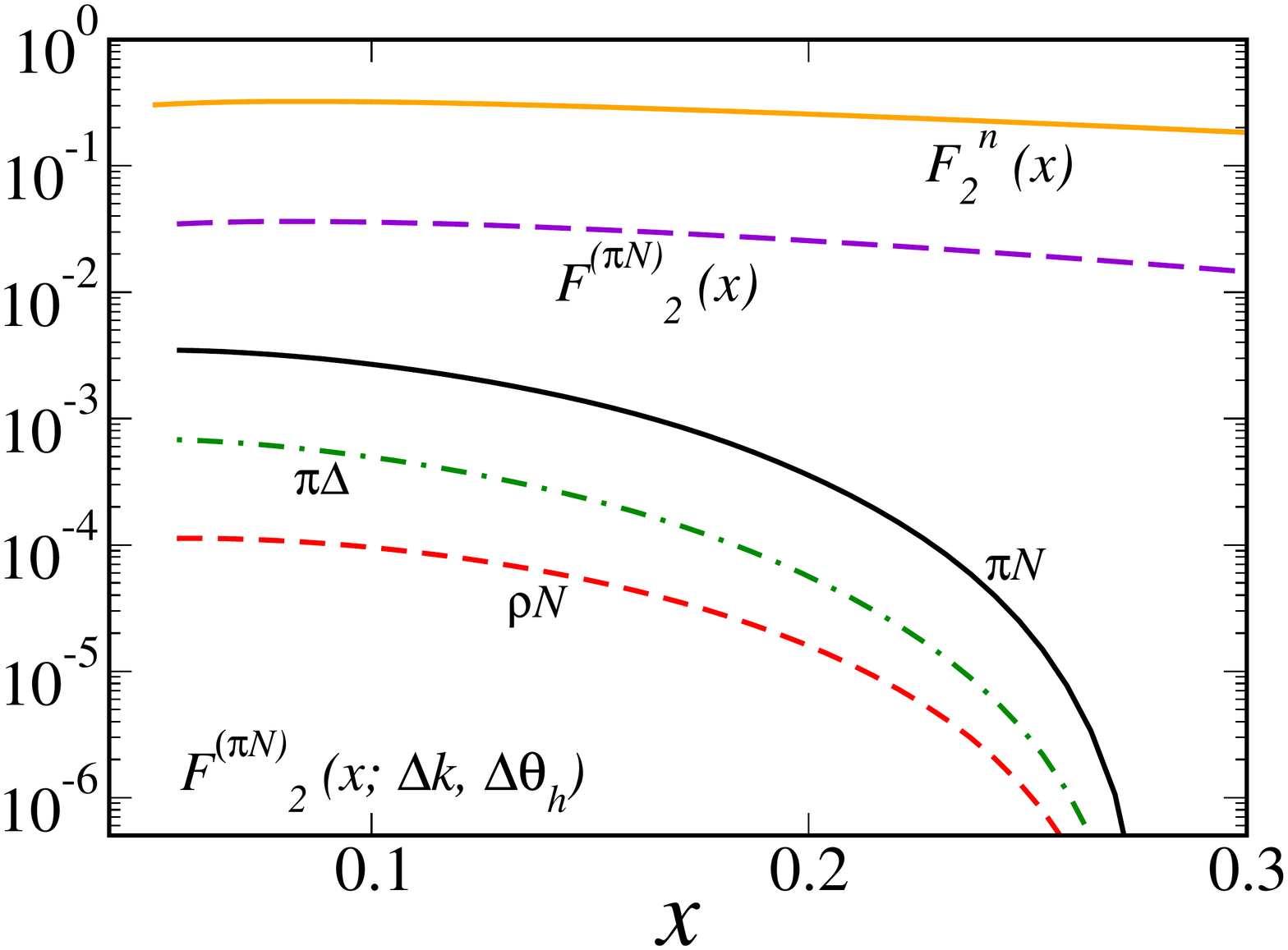} \ \ \ \
\includegraphics[width=0.48\textwidth]{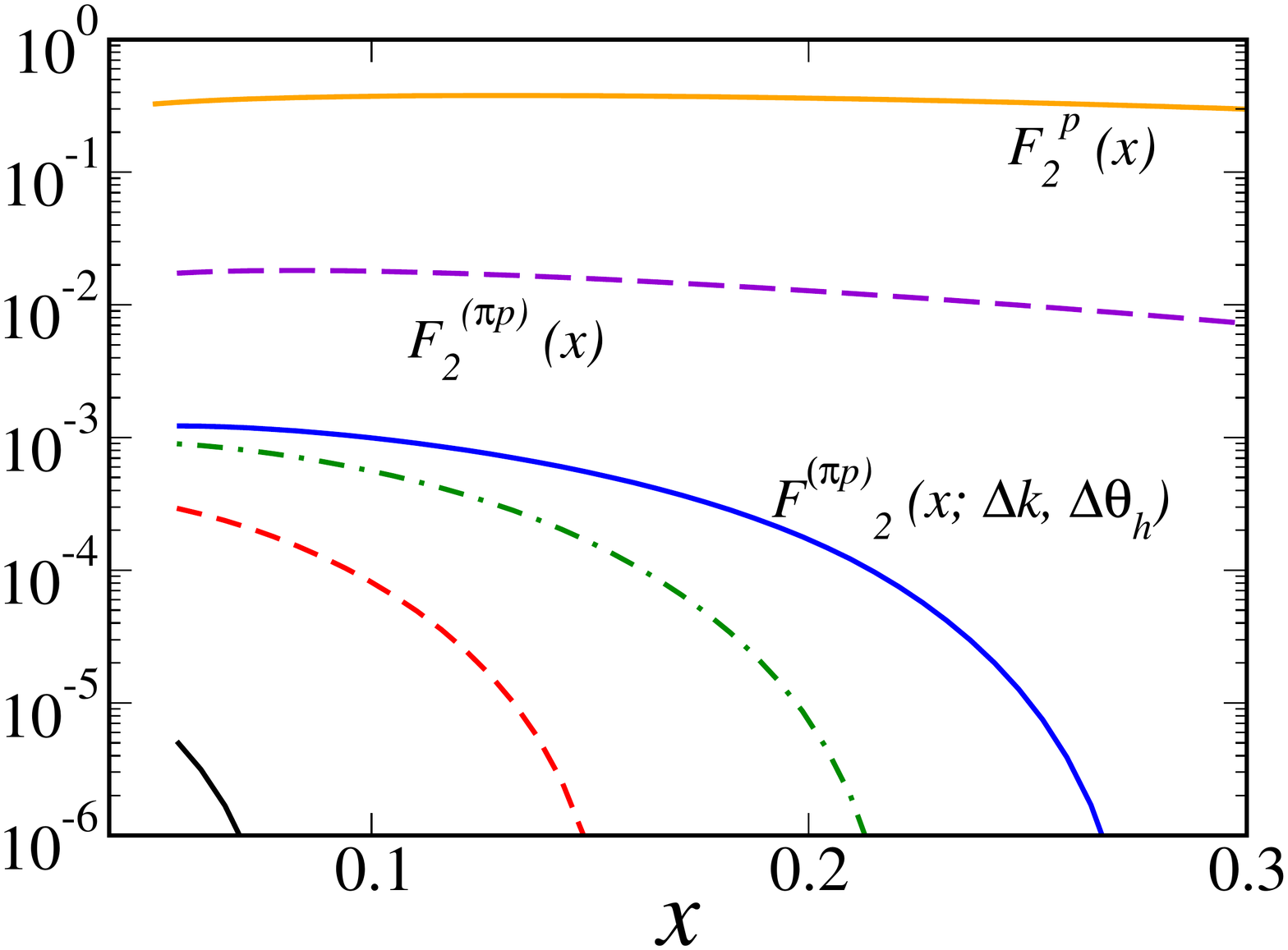}
\caption{
Contributions from tagged structure functions for various hadronic species (left), and for
various integration ranges in the final state momentum $\Delta |{\bm k}|$ (right) as determined
by Eq.~(\ref{eq:sidis_int-vc}) and explained in-text.
}
\label{fig:piFig}
\end{figure*}

Keeping with the expressions of Eq.~(\ref{eq:kin+}), it is useful to render tagged SFs in the appropriate laboratory kinematics
--- \IE~the three-momentum of the produced baryon $|{\bm k}|$ and its laboratory angle $\theta_{p'}$. Evaluating in terms
of the lab-frame definitions $\tilde{z},\ \tilde{k}_\perp$, we obtain
\begin{equation}
F_2^{(\pi N)}(x, \Delta |{\bm k}|, \Delta \theta_{p'})
= {1\over M^2} \int_{\Delta |{\bm k}|} d|{\bm k}| \int_{\Delta \theta_{p'}}d\theta_{p'}\
  J(x,|{\bm k}|,\theta_{p'})\
  f_{\pi N}(\tilde{z},\tilde{k}_\perp)\, F_{2\pi}\Big(\frac{x}{\tilde{z}}\Big)\ ,
\label{eq:sidis_int-vc}
\end{equation}
and, much as before, $\Delta |{\bm k}|, \Delta \theta_{p'}$ denote finite integration ranges determined by experimental
resolution; the quantity $J(x,|{\bm k}|,\theta_{p'})$ is the Jacobian associated with the transformation from the ``theory''
space relevant for Eq.~(\ref{eq:sidis_int}) to the lab-frame parametrization used above. Explicitly,
\begin{align}
J(x,|{\bm k}|,\theta_{p'})\ &= {\partial(x,\tilde{z},\tilde{k}^2_\perp) \over \partial(x,|{\bm k}|,\theta_{p'})} \nonumber\\
&=\ {2 \over M} |{\bm k}|^2 \sin{\theta_{p'}} \left(1 - \sin{\phi_k}\ \cos{\theta_{p'}} \right)\ .
\label{eq:Jaco}
\end{align}
where we have defined $\phi_k\ =\ \tan^{-1}(|{\bm k}|/M)$.

Using lagrangians similar to Eq.~(\ref{eq:piN-PV}), one may compute probability distributions analogous to
Eq.~(\ref{eq:piN-SF}) for $\rho$ exchange and $\Delta$ recoils --- \IE~$f_{\rho N}(z)$, $f_{\pi \Delta}(z)$ \cite{HSS}
--- and use the formalism summarized by Eqs.~(\ref{eq:conv})--(\ref{eq:Jaco}) to determine their contributions to the
tagged $F_2$ structure function. As an illustration, one might assume the experimental range of the proposed JLab Hall A
measurements described in Ref.~\cite{TDIS}. Here, the drive is to probe softer exchanges (\IE~near the $t = m^2_\pi$ pion
pole), and we therefore integrate over the modest range $\Delta |{\bm k}| = [250, 400\ {\rm MeV}]$, and include all
accessible baryon production angles in the left panel of Fig.~\ref{fig:piFig}. At such small momenta, the contributions
to $F_2^{(\pi N)}(x, \Delta |{\bm k}|, \Delta \theta_{p'})$ from single pion exchange dominate other mechanisms (which possess
more nonlinear momentum dependences) by an approximate order-of-magnitude.

The observed level of separation among the various $SU(2)$ processes in the tagged structure functions implies that a more sensitive
unraveling of the pion Sullivan process may be achievable; to that end, the detailed momentum dependence of this latter process is
depicted in the right-hand side of Fig.~\ref{fig:piFig}. Presuming a high degree of resolution in $|{\bm k}|$, the RHS of
Fig.~\ref{fig:piFig} renders the single neutral-pion exchange contributions to $F_2$ as given by Eq.~(\ref{eq:sidis_int-vc}) in fine windows
of $\Delta |{\bm k}| = [0.06, 0.1],\ [0.1-0.2],\ [0.2-0.3]$, and $[0.3-0.4\ {\rm GeV}]$ --- going respectively from the solid black to solid
blue curves. As in the left panel of Fig.~\ref{fig:piFig}, the fully integrated $\pi N$ contributions (violet, dashed --- given by Eq.~[\ref{eq:conv}])
and full DIS structure functions (either $F_2^n$ or $F_2^p$, in solid orange) are given for comparison.

In particular, there is again a fairly marked separation, this time in $\Delta |{\bm k}|$ due to momentum tagging of the forward spectator
baryon. With sensitivity to tagged SFs/fracture functions at the level of $\sim 4 \times 10^{-5}$ as might be provided by the upgraded Hall A
projection chamber \cite{TDIS}, careful measurement could in fact better elucidate the $\pi N$ vertex, whose description remains heavily
model-dependent, but nonetheless explicitly controls the Sullivan process of Fig.~\ref{fig:Feyn}. These phenomenological improvements
would of course also enhance the understanding of the pion structure function $F_2^\pi$ itself, as well as yield unprecedented information
near the pion's physical pole.
 

\section{Global analysis of intrinsic charm}
\label{sec:GA}

While the meson cloud is especially adapted to the light sector --- where chiral symmetry breaking provides physical motivation for the
pion cloud picture of nucleon structure --- Eq.~(\ref{eq:Fock}) is sufficiently general to allow an extension of the preceding formalism
to heavier mass scales. In fact, an extension of this type might serve as the basis for an investigation of the contributions to nucleon
structure due to nonperturbative heavy quarks, including {\it intrinsic} charm (IC) \cite{BHPS}.

In Ref.~\cite{Hobbs:2013bia}, precisely such an analysis was explored, with the end result being a two-step calculation yielding non-zero
IC PDFs of the form 
\begin{equation}
c(x) = \sum_{B,M}\,
    \Big[ \int_x^1 \frac{d\bar z}{\bar z}\,
          f_{BM}(\bar z)\, c_B\Big(\frac{x}{\bar z}\Big)
    \Big]\ , \ \ \
\bar{c}(x) = \sum_{M,B}\,
    \Big[ \int_x^1 \frac{dz}{z}\,
          f_{MB}(z)\, \bar{c}_M\Big(\frac{x}{z}\Big) \Big]\ ,
\label{eq:IC_conv}
\end{equation}
where $\bar{z} = 1-z$, and $z$ retains its definition as given in Sec.~\ref{sec:TDIS}. In contrast to the foregoing analysis,
the sums over $(B, M)$ of Eq.~(\ref{eq:IC_conv}) now involve (anti-)charm-containing $SU(4)$ hadrons, and the hadronic splitting
functions $f_{MB}(z)$ must be evaluated using a hadronic effective theory. It is of course also necessary to determine the intermediate
distributions $c_B(x/\bar{z})$ and $\bar{c}_M(x/z)$, which were computed using a simple relativistic quark model formulated on the
light front as detailed in \cite{Hobbs:2013bia}.

With our framework, one may then incorporate intrinsic charm at the partonic threshold $Q^2 = m^2_c \sim 1.69$ GeV$^2$ and
investigate implications for higher energy data using standard QCD evolution and the various IC model prescriptions treated
in \cite{Hobbs:2013bia}; doing so, it was found that hadroproduction data might serve to constrain the model
parameters that enter regulators analogous to $G_{\pi N}$, but that the resulting predictions for $F_2^{c\bar{c}}$ at the
upper reaches of the $Q^2$ of measurements recorded by the European Muon Collaboration (EMC) \cite{EMC} were rather large. As these measurements have
been cited as evidence of IC \cite{Brod-Evid}, a more systematic analysis of the model developed in \cite{Hobbs:2013bia} using the
technology of a QCD global fit is especially timely. Moreover, a thorough global analysis would also be capable of assessing
the impact of the suggestive EMC data in a setting that comprehensively allows IC.

Such an effort has recently been carried out and described in Ref.~\cite{GlobA}; for this purpose, it was convenient to fit the models in
\cite{Hobbs:2013bia} to simple, three-parameter expressions of the form
\begin{equation}
c(x) = C^{(0)}\, A\, x^\alpha (1-x)^\beta\ , \hspace*{1cm}
\bar{c}(x) = C^{(0)}\, \bar{A}\, x^{\bar\alpha} (1-x)^{\bar\beta}\ ,
\label{eq:3-par}
\end{equation}
where the normalization constants are defined in terms of beta functions as
  $A = 1/B(\alpha+1,\beta+1)$ and
  $\bar{A} = 1/B(\bar\alpha+1,\bar\beta+1)$,
so as to enforce the normalization of the distributions to $C^{(0)}$.
The overall normalization $C^{(0)}$ in Eq.~(\ref{eq:3-par}) is then directly related to the total momentum carried by IC, given by
\begin{equation}
\langle x \rangle_{\rm IC} = \int_0^1\ dx\ x\  \{c + \bar{c} \}(x)\ ;
\end{equation}
and one can collectively fit the world data involving various reactions over a wide kinematical range.

The QCD global fit was performed using the JR14 framework developed in Ref.~\cite{JR14}, which among other things provides a
sensitive treatment of finite-$Q^2$ corrections due to higher twist and target/quark mass effects. We are thus enabled to
incorporate data with less restrictive cuts in $W^2$ and $Q^2$ as compared with previous efforts, \EG~\cite{CT}, that allowed
a significant IC component to the nucleon. For the sake of the analysis, the $F_2$ structure function is assumed to be a
combination of light and heavy contributions $F_2 = F_2^{\rm light} + F_2^{\rm heavy}$,
in which $F_2^{\rm light}$ represents the ($u$, $d$, $s$) light-sector piece, whereas $F_2^{\rm heavy}$ contains
the heavy $c$ and $b$ quarks. The charm-sector component of $F_2^{\rm heavy}$ may be further broken into a
part which is generated through pQCD-calculable processes, $F_2^{c\bar c}$, as well as an intrinsic part, $F_2^{\rm IC}$;
\IE~$F_2^c = F_2^{c\bar c} + F_2^{^{\rm IC}}$.
The pQCD component $F_2^{c\bar{c}}$ at intermediate $Q^2$ is described by the photon-gluon fusion mechanism
computed formally in Ref.~\cite{HM}; in particular, calculating the necessary diagrams to NLO in $\alpha_s$
yields
\begin{equation}
F_2^{c\bar c}(x,Q^2,m_c^2)
= \frac{Q^2 \alpha_s}{4\pi^2 m_c^2}
  \sum_i \int\frac{dz}{z}\
  \hat\sigma_i(\eta,\xi)\ f_i\Big(\frac{x}{z}, \mu\Big)\ ,
\label{eq:F2cc}
\end{equation}
with $\hat\sigma_i$ denoting the hard, parton-level scattering of
a flavor $i = u, d, s$ or $g$ into final states consisting of a $c\bar c$
pair. On the other hand, $f_i$ represents the associated parton densities in the nucleon.

The Hoffmann-Moore scheme of \cite{HM} systematically incorporates mass-dependent threshold
effects, and $\hat\sigma_i$ must therefore be computed in terms of the parameters
$\xi = Q^2/m_c^2$ and $\eta = Q^2 \cdot (1-z)\big/(4m^2_c z) - 1$; additionally,
the parton densities of Eq.~(\ref{eq:F2cc}) must also be evaluated at a factorization scale specified
by $\mu_F^2 = 4 m_c^2 + Q^2$, in which we take the charm mass to be $m_c = 1.3$~GeV at threshold
for consistency with the `confining' prescription of \cite{Hobbs:2013bia}. We point out also that
the present analysis assumed a partonic threshold $W^2 > (2 m_c)^2$ in evaluating $F^c_2$. Relating
this condition to the correspondingly larger hadronic threshold for the production of charm-containing final
states is generally model-dependent, but has been undertaken in Ref.~\cite{GlobA}. It should be
noted that hadronic threshold effects can potentially impose more restrictive kinematical cuts
on the data involved, and therefore slightly weaken the sharp growths in $\chi^2$ presented below.

\begin{figure*}
\centering
\includegraphics[width=0.48\textwidth]{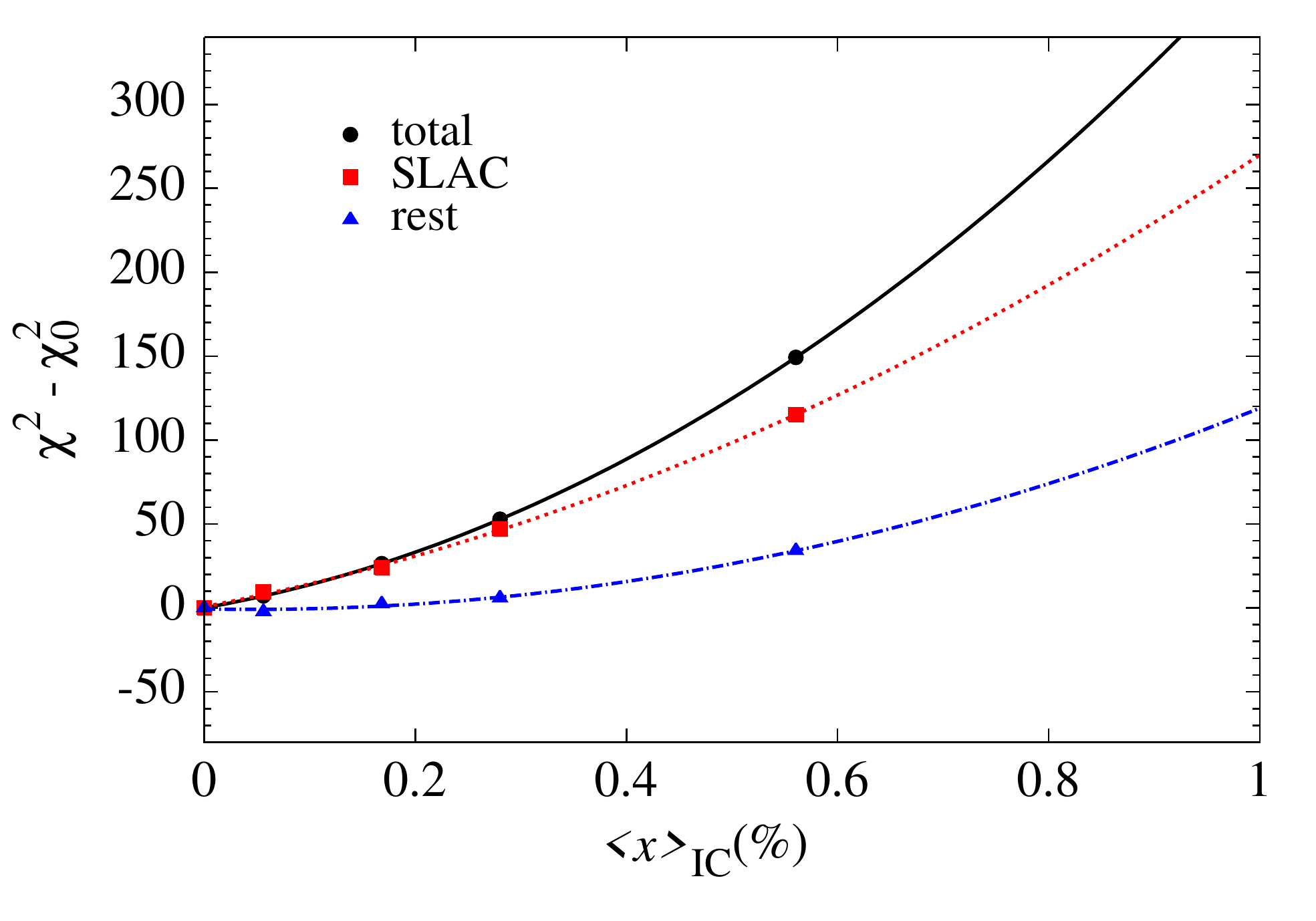} \ \ \ \ 
\includegraphics[width=0.48\textwidth]{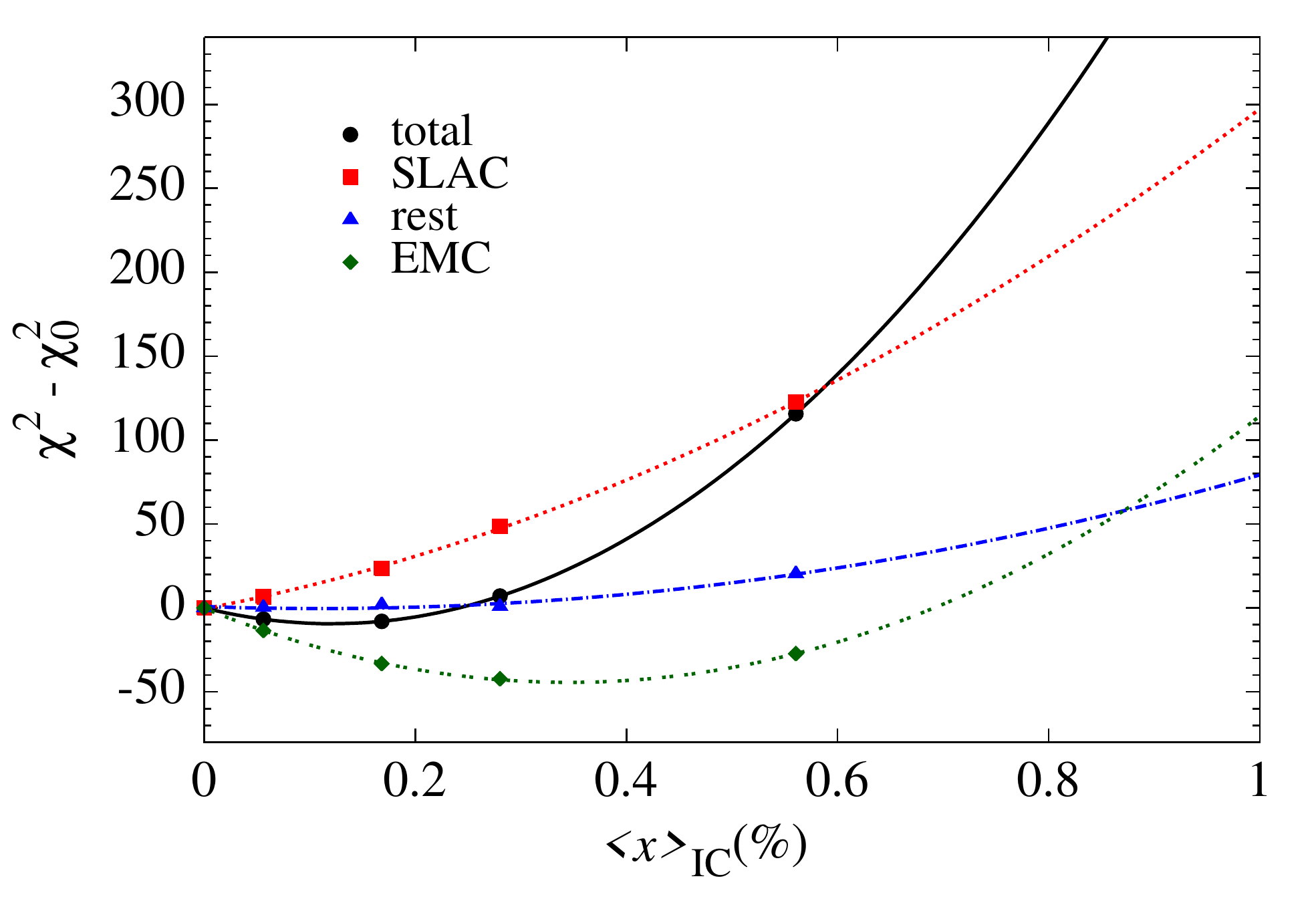}
\caption{
Dependence on $\langle x \rangle_{\rm IC}$ of the $\chi^2$ of fits that omit (left) and
incorporate (right) the EMC $F_2^{c\bar{c}}$ data.
}
\label{fig:GAFig}
\end{figure*}

Of the data we incorporate at lower $Q^2$ and $W^2$, $ep$ and $ed$ fixed-target DIS measurements \cite{SLAC}
from SLAC are especially influential in limiting the magnitude of $\lan x \ran_{\rm IC}$, as we explore in
Fig.~\ref{fig:GAFig}. In both panels of Fig.~\ref{fig:GAFig}, we plot the rise in $\chi^2$ against the
magnitude of $\lan x \ran_{\rm IC}$ assumed in the IC model parametrized according to Eq.~(\ref{eq:3-par}).
Moreover, we globally fit data in this fashion twice --- excluding and actively including the EMC extractions of $F_2^c$
as shown in the left and right panels of Fig.~\ref{fig:GAFig}, respectively.
The results of this procedure are striking: the rapid takeoff of $\chi^2$ from $\lan x \ran_{\rm IC} \sim 0$
indicates plainly that the world data sharply limit the size of IC, with the SLAC measurements serving
as a principal constraint (as illustrated by the `scans,' which explicitly separate the contributions
to the growth of $\chi^2$). In fact, the `total' fit of the LHS of Fig.~\ref{fig:GAFig} amounts to a
5$\sigma$ limit of $\lan x \ran_{\rm IC} \le 0.1\%$.

In the end, inclusion of the EMC points of Ref.~\cite{EMC} does little to soften this restrictive bound ---
while the scan corresponding to the EMC set (green dot-dashed) in the right panel of Fig.~\ref{fig:GAFig}
exhibits a minimum in $\chi^2$ about $\lan x \ran_{\rm IC} \sim 0.3 - 0.4\%$, this shallows significantly
when combined with the global data; the `total' fit in this case prefers $\lan x \ran_{\rm IC} = 0.13 \pm 0.04\%$.
In Ref.~\cite{GlobA} we noted that the EMC data themselves exhibit some degree of tension with other segments
of the data sets we fit, most notably, HERA measurements of the reduced charm cross section $\sigma^{c\bar{c}}_r$ \cite{HERA}.
Specifically, these data occupy partially overlapping regions of the kinematical parameter space with respect to
the older EMC measurements, but are comparably much better fit by the global analysis (unlike the EMC points, which
have $\chi^2 / n_{d.o.f.} \sim 4.3$). Evidence of such tension in data that weigh critically in favor of IC should
do nothing but urge the necessity of new and improved measurements of charm production, especially at large $x$.

\begin{acknowledgements}
The author acknowledges J.~T.~Londergan, W.~Melnitchouk, C.~E.~Keppel, B.~Wojtsekhowski, and
P.~Jimenez-Delgado for collaboration on projects related to this note, as well as G.~A.~Miller for helpful discussions.
The author also gratefully acknowledges support from the Gary T.~McCartor Fund Fellowship.
\end{acknowledgements}


\end{document}